\documentclass[aps,eqsecnum,nofootinbib,showpacs,twocolumn]{revtex4}
\usepackage{amsmath,amsfonts,amssymb,bm}
\usepackage[dvips]{graphicx}
\usepackage{color}
\usepackage{wrapfig}
\usepackage{graphicx}
\usepackage{color}  
\def\be{\begin{eqnarray}}
\def\ee{\end{eqnarray}}
\def\bec{\begin{center}}
\def\eec{\end{center}}

\def\p{\partial}
\bmdefine{\bmk}{\bm{k}}
\bmdefine{\bmx}{\bm{x}}
\newcommand{\hS}{\Hat{S}}
\newcommand{\hrho}{\Hat{\rho}}

\newcommand{\calI}{\mathcal{I}}

\newcommand{\calL}{\mathcal{L}}
\newcommand{\calM}{\mathcal{M}}
\newcommand{\scp}[1]{{(#1)}}
\newcommand{\odiff}[2]{ \frac{d #1}{d #2} }
\newcommand{\odiffII}[2]{ \frac{d^2 #1}{d #2^2} }
\newcommand{\pdiff}[2]{ \frac{\partial #1}{\partial #2} }
\newcommand{\pdiffII}[2]{ \frac{\partial^2 #1}{\partial #2^2} }

\newcommand{\IP}[2]{\langle~#1~\vert~#2~\rangle}
\newcommand{\Exp}[1]{\left\langle~#1~\right\rangle}

\begin{document}
\title{Vortex lattice for a holographic superconductor}
\author{Kengo Maeda}
\email{maeda302@sic.shibaura-it.ac.jp}
\affiliation{Faculty of Engineering,
Shibaura Institute of Technology, Saitama, 330-8570, Japan}

\author{Makoto Natsuume}
\email{makoto.natsuume@kek.jp}
\affiliation{KEK Theory Center, Institute of Particle and Nuclear Studies, 
High Energy Accelerator Research Organization,
Tsukuba, Ibaraki, 305-0801, Japan}

\author{Takashi Okamura}
\email{tokamura@kwansei.ac.jp}
\affiliation{Department of Physics, Kwansei Gakuin University,
Sanda, 669-1337, Japan}

\date{\today}
\begin{abstract}
We investigate the vortex lattice solution in a $(2+1)$-dimensional holographic model of superconductors constructed from a charged scalar condensate. The solution is obtained perturbatively near the second-order phase transition and is a holographic realization of the Abrikosov lattice. Below a critical value of magnetic field, the solution has a lower free energy than the normal state. Both the free energy density and the superconducting current are expressed by nonlocal functions, but they reduce to the expressions in the Ginzburg-Landau (GL) theory at long wavelength. As a result, a triangular lattice becomes the most favorable solution thermodynamically as in the GL theory of type II superconductors. 
\end{abstract}


\maketitle

\section{Introduction}\label{sec:intro}

The application of AdS/CFT (anti-de Sitter/conformal field theory)
correspondence to condensed matter physics has recently become
one of the most interesting topics in string theory. (See Refs.~\cite{Hartnoll:2009sz,Herzog:2009xv} for reviews.) 
In Refs.~\cite{gubser77,HHH2008},
a holographic model of a superconductor is constructed
by a gravitational theory of a complex scalar field
coupled to the Maxwell field.
This opens up a window to study superconductors
in strongly interacting system via AdS/CFT correspondence.

The magnetic property of the superconductors, 
such as the Meissner effect, is a good place to start the investigation 
since it plays the crucial role to distinguish between type I and type II
superconductors \cite{parks}.
Recently, many localized solutions under the external magnetic field $B$ were constructed numerically, which include a ``droplet'' solution
and a single vortex solution with integer winding number
\cite{nakanowen,albashjohnson2008,albashjohnson2009verI,
albashjohnson2009verII,MPS,Keranen:2009vi,Keranen:2009ss}.
These numerical solutions imply that the holographic superconductor belongs to type II, in agreement with the scaling argument in Ref.~\cite{HHH2008verII}.

There are two critical magnetic fields $B_{c1}$ 
(the lower critical value) and $B_{c2}$ (the upper critical value)
in type II superconductors.
At $B=B_{c1}$, the external magnetic field begins to penetrate
into the superconductor and vortices appear for $B>B_{c1}$.
At $B=B_{c2}$, the second-order phase transition occurs and
the superconductivity disappears.
When $B$ approaches $B_{c2}$ from below, a vortex lattice,
in which a single vortex is arranged periodically, should appear
since it is  more favorable thermodynamically
than a single vortex.

In this paper, we construct the vortex lattice solution or the Abrikosov lattice solution, which is characterized by two lattice parameters, $a_1$ and $a_2$,
perturbatively near the second-order phase transition.
We follow the treatment of type II superconductors
based on Ginzburg-Landau~(GL) equations.
The solution includes a triangular lattice solution,
which is known as the most favorable solution thermodynamically
in the GL theory.
We determine the critical value $B_{c2}$
and obtain the free energy parametrized by $a_1$ and $a_2$.
It is shown that the free energy is always smaller than that of normal state for any choice of two parameters.

In the conventional Abrikosov lattice, there exists a circulating superconducting current around the core, and the current flows along the lines of constant field for the condensate [See Eq.~(\ref{GL-current})]. We evaluate the R-current to see if the current in the holographic superconductor has the similar property. 

It turns out that both the free energy density and the R-current are written by nonlocal functions of the condensate unlike the GL theory. 
However, these expressions reduce to the ones in the GL theory at long wavelength.

The plan of our paper is as follows:
In Sec.~\ref{sec:solution}, we construct the vortex lattice solution
by superposing the single droplet solutions
found in Ref.~\cite{albashjohnson2008}.
We obtain the upper critical magnetic field $B_{c2}$.
In Sec.~\ref{sec:free_energy}, we calculate the free energy and the R-current 
in a power series expansion in the order parameter.
In Sec.~\ref{sec:low_energy_limit}, we take the long-wavelength limit. In this case, 
the triangular lattice solution minimizes the free energy, and the R-current flows 
along the lines of constant field for the condensate.
Section~\ref{sec:discussion} is devoted to conclusion and discussion.

\section{Phase diagram and solutions}\label{sec:solution}

We consider a (2+1)-dimensional holographic superconductor
described by a dual gravitational theory in four dimensions~($AdS_4$)
coupled to a charged complex scalar field $\Psi$
and a Maxwell field $A_\mu$~\cite{HHH2008}.
To investigate the superconducting phase
near the second-order phase transition,
we consider the equations of motion close to the phase transition,
as considered in Refs.~\cite{maedaokamura2008,parks}.
For simplicity, we take a probe limit
where the backreaction of the matter field onto the geometry
can be ignored \cite{HHH2008}.

In this section, we first obtain a ``droplet'' solution similar to the one obtained in Ref.~\cite{albashjohnson2008} by solving the equations of motion at the leading order.
We also obtain the upper critical magnetic field $B_{c2}$
as a function of $T$.
Then, we construct the vortex lattice solution
by superposing the droplet solutions.

\subsection{``Droplet'' solution}

The background metric is given by $AdS_4$-Schwarzschild black hole
with metric
\begin{subequations}
\begin{align}
  & ds^2 = \frac{L^2 \alpha^2}{u^2} ( - h(u) dt^2 + dx^2 + dy^2 )
  + \frac{L^2 du^2}{u^2 h(u)},
\label{eq:line_element} \\
  & h(u) = 1 - u^3,
  \hspace{1.0truecm}
  \alpha(T) = \frac{4 \pi T}{3} = \frac{R_0}{L^2},
\label{eq:def-h_alpha}
\end{align}
\label{metric}%
\end{subequations}
where $L$, $R_0$, and $T$ are AdS radius, horizon radius,
and the Hawking temperature, respectively.
We take the coordinate $u$
such that the horizon is located at $u=1$.
The action of the matter system
$S = (L^2/2 \kappa_4^2 e^2) \hS$
is written by
\begin{align}
  & \hS
  = \int_{\calM} d^4x~\sqrt{-g} \left(
  - \frac{F^2}{4} - |D \Psi |^2 - m^2 | \Psi |^2 \right),
\label{action1}
\end{align}
where $m$ and $e$ are the mass and charge of the scalar field $\Psi$,
respectively,
and
\begin{align}
  & D_\mu = \nabla_\mu - i A_\mu,
  \hspace{1.0truecm}
  F_{\mu\nu} = \p_\mu A_\nu - \p_\nu A_\mu.
\label{eq:def-cov_deri_D}
\end{align}

The probe limit is realized by taking the limit $e \to \infty$,
keeping $A_\mu$ and $\Psi$ fixed.
The equations of motion are given by
\begin{subequations}
\begin{align}
  & D^2 \Psi - m^2 \Psi = 0,
\label{eq:EOM-Psi} \\
  & \nabla_\nu F_\mu{}^\nu = j_\mu
  := i [ (D_\mu\Psi)^\dagger \Psi - \Psi^\dagger (D_\mu\Psi) ].
\label{eq:EOM-A}
\end{align}
\label{eq-motion}%
\end{subequations}
Hereafter, we choose a gauge $A_u=0$. We will consider stationary solutions since our interest is in thermodynamics of the holographic superconductor. Then, 
Eqs.~(\ref{eq-motion}) become
\begin{subequations}
\begin{align}
  & \left( u^2 \pdiff{}{u} \frac{h}{u^2} \pdiff{}{u}
  + \frac{A_t^2}{\alpha^2 h} - \frac{m^2 L^2}{u^2} \right) \Psi
\nonumber \\
  &\hspace{0.5truecm}
  = - \frac{1}{\alpha^2}\, \delta^{ij} D_i D_j \Psi,
\label{eq:EOM-Psi_gf} \\
  & \left( \alpha^2 h\, \pdiffII{}{u} + \triangle \right) A_t
  = \frac{2 L^2 \alpha^2}{u^2} A_t | \Psi |^2,
\label{eq:EOM-At_gf} \\
  & \left( \pdiff{}{u}\, \alpha^2 h\, \pdiff{}{u}
  + \triangle \right) A_{i}
  - \partial_i \left( \delta^{jk} \partial_j A_{k} \right)
\nonumber \\
  &\hspace{0.5truecm}
  = - \frac{L^2 \alpha^2}{u^2}\, j_i
  ,
\label{eq:EOM-Ai_gf} \\
  & \partial_u \left( \delta^{ij} \partial_i A_{j} \right)
  = \frac{2 L^2 \alpha^2}{u^2}
    \Im\left( \Psi^\dagger \partial_u \Psi \right) ,
\label{eq:div-Ai_gf}
\end{align}
\label{eq:EOM-A_zeroth}%
\end{subequations}
where $i, j, k = x, y$ and $\triangle = \p_x^2 + \p_y^2$.

We solve these equations under the following boundary conditions:
\begin{itemize}
\item For the scalar field, we shall confine our interest to $m^2L^2=-2$ case. Then, the asymptotic behavior of $\Psi$ is
\begin{equation*}
\Psi \sim c_1 u + c_2 u^2 \qquad (u \to 0),
\label{b.c.-scalar}
\end{equation*}
and both modes are normalizable \cite{KlebanovWitten}. In this case, both $c_1$ and $c_2$ can be interpreted as expectation values of the dual operators with $\Delta=1$ and $\Delta=2$, respectively \cite{HHH2008}. For simplicity, we consider only the case where the faster falloff is dual to the expectation value, i.e., $c_1=0$. We also impose that $\Psi$ is regular at the horizon $u=1$.

\item The asymptotic values of the gauge field give the chemical potential $\mu$ and the external magnetic field $B$:
\begin{equation*}
\mu=A_t(\bmx, u=0), \quad B(\bmx) = F_{xy}(\bmx,u=0).
\end{equation*}
The boundary condition at the horizon is given by requiring that $A_\mu dx^\mu$ has a finite norm there, i.e., $A_i(\bmx,u=1)$ is regular and $A_t(\bmx,u=1)=0$.
\end{itemize}
We will only change the external magnetic field $B$
perpendicular to the AdS boundary,
keeping the temperature $T$ and and the chemical potential $\mu$
on the boundary theory fixed.
In this case, the scalar field $\Psi$ begins to condensate
below a critical value of the magnetic field $B_{c2}$,
while the condensation does not occur above the critical value.

Defining the deviation parameter $\epsilon$
from the critical magnetic field $B_{c2}~(>0)$ as
$\epsilon:=(B_{c2}-B)/B_{c2}$, we can expand the scalar field $\Psi$,
the gauge field, and the current $j_\mu$
as a series in $\epsilon$
\footnote{This type of expansion has been applied to
the SU(2) model of the holographic superconductor and
the speed of second sound was analytically derived~\cite{herzog-pufu}.}:
\begin{subequations}
\label{expansion}
\begin{align}
  & \Psi({\bm x},u)
  = \epsilon^{1/2}\psi_1({\bm x},u) + \epsilon^{3/2}\psi_2({\bm x},u)
  + \cdots,
\\
  & A_\mu({\bm x},u)
  = A^{(0)}_\mu({\bm x},u) + \epsilon A^{(1)}_\mu({\bm x},u) + \cdots,
\\
  & j_\mu({\bm x},u)
  = \epsilon j^{(1)}_\mu({\bm x},u) + \epsilon^2 j^{(2)}_\mu({\bm x},u)
  + \cdots,
\end{align}
\end{subequations}
where $\bmx=(x,y)$.

The zeroth order solution generating the critical homogeneous
magnetic field $B_{c2}$~$(>0)$
and the chemical potential $\mu$ are given by
\begin{align}
  & A^\scp{0}_t = \mu (1-u),
& & A^\scp{0}_x = 0,
& & A^\scp{0}_y = B_{c2} x.
\label{sol-A-zeroth}
\end{align}
%
\begin{figure}
\includegraphics[width=8.0truecm,clip]{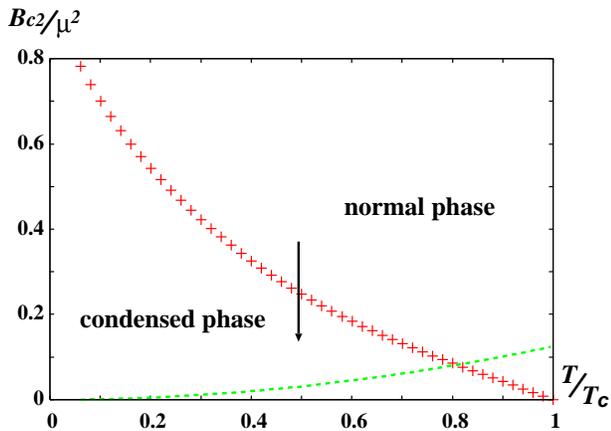}
\caption{ \label{fig:phase_diagram}
(color online).
$B_{c2}$($+$) as a function of $T/T_c$.
Dashed line: $B/\mu^2 = (1/8) (T/T_c)^2$ (See Sec.~\ref{sec:low_energy_limit}).
}
\end{figure}

Substituting Eq.~(\ref{sol-A-zeroth})
into Eq.~(\ref{eq:EOM-Psi_gf})
and taking an ansatz
$\psi_1(\bmx,u)=e^{ipy}\phi(x,u;p)/L$ for a constant $p$,
the equation of motion for $\psi_1$ reduces to
\begin{align}
  & \left[~u^2 \pdiff{}{u} \left( \frac{h(u)}{u^2} \pdiff{}{u} \right)
  + \frac{\big( A^\scp{0}_t(u) \big)^2}{\alpha^2 h(u)}
  - \frac{m^2 L^2}{u^2}~\right] \phi(x,u;p)
\nonumber \\
  =&~\frac{1}{\alpha^2} \left[ - \pdiffII{}{x}
  + \left( p - B_{c2} x \right)^2~\right] \phi(x,u;p) .
\label{eq-psi-first}
\end{align}

Following the ansatz in Ref.~\cite{albashjohnson2008},
we separate the variable $\phi$ as
$\phi_n(x,u; p)=\rho_n(u)\gamma_n(x;p)/L$
with a separation constant $\lambda_n$.
The equations for $\rho_n$ and $\gamma_n$ are divided
into the following equations:
\begin{subequations}
\begin{align}
  & \left( - \pdiffII{}{X} + \frac{X^2}{4} \right)
  \gamma_n(x;p)
  = \frac{\lambda_n}{2}\, \gamma_n(x;p) ,
\label{eq-gamma} \\
  & h \rho_n''(u) - \left( \frac{2 h}{u} + 3 u^2 \right) \rho_n'(u)
\nonumber \\
  &\hspace*{0.5truecm}
  = \left( \frac{m^2 L^2}{u^2}
  - \frac{q^2}{h} (1-u)^2
  + q^2 \frac{B_{c2} \lambda_n}{\mu^2}
  \right) \rho_n ,
\label{eq-rho}
\end{align}
\end{subequations}
where $X := \sqrt{2 B_{c2}} ( x - p/B_{c2} )$ and $q:=\mu/\alpha$
is a dimensionless quantity.

The regular and bounded solution of Eq.~(\ref{eq-gamma}) satisfying
$\lim_{|x|\to\infty }|\gamma_n|<\infty$
is given by Hermite functions $H_n$ as
\begin{align}
  & \gamma_n(x;p) = e^{- X^2/4 } H_n(X),
\end{align}
and the corresponding eigenvalue $\lambda_n$ is
\begin{align}
  & \lambda_n = 2 n + 1 ,
\end{align}
for a non-negative integer $n$.

This solution corresponds to a ``droplet'' solution 
obtained in a earlier work~\cite{albashjohnson2008}
in the sense that these solutions fall off rapidly at large $|x|$. 
A single (localized) droplet solution is easily obtained
by generalizing the solution
with circular symmetry, as shown
in Refs.~\cite{albashjohnson2009verI,albashjohnson2009verII,MPS}.

Now, let us consider the phase diagram. 
Equation~(\ref{eq-rho}) is characterized by two parameters $T/\mu \propto 1/q$ and $B_{c2} \lambda_n/\mu^2$. Also, recall that we impose boundary conditions both at asymptotic infinity and at the horizon. 
Such a problem has a nontrivial solution only when there is a relation between these two set of parameters. The solution corresponds to the case where the  $\Psi^\scp{0} = 0$ state becomes marginally stable. 


%

If one has a second-order phase transition from the $\Psi = 0$ state to the ``hairy" black hole, one should encounter a marginally stable state at the point of transition. So, the solution of Eq.~(\ref{eq-rho}) gives the candidate for the phase transition point.

It is clear which side of the phase transition line is the superconducting phase. 
Suppose that one lowers the magnetic field with a fixed $T/\mu$. The right-hand side of Eq.~(\ref{eq-rho}) suggests that the magnetic field increases $(\text{effective mass})^2$ and tends to stabilize the state. [On the other hand, the electric field decreases $(\text{effective mass})^2$ and tends to destabilize the $\Psi^\scp{0} = 0$ state.] 
Thus, the $\Psi^\scp{0} = 0$ state should be stable under a large enough magnetic field, and the state becomes unstable as one lowers the magnetic field. 
Then, it is likely that the upper critical magnetic field is given by the value of the magnetic field when one first encounters a marginally stable state (as one lowers the magnetic field). Note that a marginally stable solution is parametrized by $B \lambda_n/\mu^2$ for a fixed $T/\mu$. Thus, one has the largest magnetic field when $\lambda_n$ takes the minimum, namely the $n=0$ solution. 

Consequently, the upper critical magnetic field $B_{c2}$ is given by the largest $B/\mu^2$ for the solution of Eq.~(\ref{eq-rho}). Figure~\ref{fig:phase_diagram} shows $B_{c2}$ obtained in this manner in the $(T/T_c, B/\mu^2)$ phase diagram.%
\footnote{Obviously, the above argument gives only the necessary condition for the phase transition and does not give the sufficient condition. The sufficient condition is given by showing that there exists a condensate solution for $B < B_{c2}$ and that its free energy is lower than the $\Psi = 0$ solution. This is shown perturbatively below.}
Here, $T_c$ is the critical temperature when there is no magnetic field,
and it is determined in the combination
$T_c/\mu = 3/(4 \pi q_c)$ with
$q_c \sim 4.07$ \cite{maedaokamura2008}.

\subsection{Vortex lattice solution}

Let us clarify how the vortex lattice constructed in this subsection differs from the droplet in the previous subsection. A characteristic feature of a vortex is that it has a zero and has a winding number around the zero. Thus, the droplet which is nonvanishing everywhere cannot have such a winding number. However, a superposition of droplets can have zeros and winding numbers as we will see in a moment. 

As we have seen in the last subsection, it is enough to consider only the $n=0$ solution near $B_{c2}$:
\begin{align}
  & \gamma_0(x;p)
  = e^{- X^2/4}
  = \exp\left[ - \frac{1}{2}
    \left( \frac{x}{r_0} - p r_0 \right)^2 \right],
\label{sol-droplet}
\end{align}
where
$r_0 := 1/\sqrt{B_{c2}}$.
As $\lambda_n$ is independent of $p$,
a linear superposition of the solutions $e^{ipy}\rho_n(u)\gamma_n(x;p)$
with different $p$ is also a solution of the equation of motion
for $\Psi$ at $O(\epsilon^{1/2})$.
To obtain the vortex lattice solution
from the single droplet solution 
(\ref{sol-droplet}), consider the following superposition:
\begin{subequations}
\begin{align}
  & \psi_1({\bm x},u)
  = \frac{\rho_0(u)}{L}
  \sum_{l=-\infty}^{\infty} c_l\, e^{i p_l y} \gamma_0(x; p_l) ,
\label{eq:lattice_sol} \\
  & c_l
  := \exp\left( - i \frac{\pi a_2}{a_1^2} l^2 \right),
  \hspace{1.0truecm}
   p_l := \frac{2 \pi l}{a_1 r_0},
\label{eq:def-C_l-p_l}
\end{align}
\label{superpose-psi}%
\end{subequations}
\noindent
for arbitrary parameters $a_1$ and $a_2$.
In terms of the elliptic theta function $\vartheta_3$ defined by
\begin{align}
  & \vartheta_3(v,\tau):=\sum_{l=-\infty}^\infty q^{l^2}z^{2l}
& & ( q:=e^{\tau\pi i},~z:=e^{i\pi v} ),
\label{def-theta}
\end{align}
the summation over $l$ in Eq.~(\ref{eq:lattice_sol}) is expressed by
\begin{align}
  & \gamma_L(\bm{x})
  := \sum_{l=-\infty}^{\infty} c_l\, e^{i p_l y} \gamma_0(x; p_l)
  = e^{- x^2/2r_0^2}~\vartheta_3(v,\tau) ,
\label{def-gamma0}
\end{align}
where
\begin{align}
  & v:=\frac{-ix+y}{a_1 r_0},
& & \tau:=\frac{2\pi i-a_2}{a_1^2}.
\end{align}

The solution (\ref{superpose-psi}) or (\ref{def-gamma0})
represents a vortex lattice. The elliptic theta function $\vartheta_3$ has two properties which are useful to see the vortex lattice structure. 
First, $\vartheta_3$ has a pseudo-periodicity
\begin{subequations}
\begin{align}
  & \vartheta_3(v+1, \tau)
  = \vartheta_3(v, \tau),
\\
  & \vartheta_3(v+\tau, \tau)
  = e^{- 2 \pi i ( v + \tau/2 ) }\, \vartheta_3(v,\tau),
\end{align}
\label{pseudo-periodicity1}%
\end{subequations}
so $\psi_1$ also has a pseudo-periodicity
\begin{subequations}
\begin{align}
  & \psi_1(x,\,y,\,u)
  = \psi_1(x, y+a_1\,r_0,\,u),
\\
  & \psi_1\left(x+\frac{2\pi r_0}{a_1}, y+\frac{a_2\,r_0}{a_1},\,u\right)
\nonumber \\
  &\hspace{0.5truecm}
  = \exp\left[ \frac{2 \pi i}{a_1}
  \left( \frac{y}{r_0} + \frac{a_2}{2 a_1} \right) \right]
  \psi_1(x,\,y,\,u).
\end{align}
\label{pseudo-periodicity2}%
\end{subequations}
Thus,
$\sigma(\bm{x}) := |\gamma_L(\bm{x})|^2$
represents a lattice
in which the fundamental region $V_0$ is spanned by two vectors
${\bm b}_1=a_1 r_0\p_y$ and
${\bm b}_2=2\pi r_0/a_1\p_x+a_2 r_0/a_1\p_y$
and the area is given by $2\pi r_0^2$.
This is the well-known result,
where the magnetic flux penetrating the unit cell is given by
$B_{c2} \times (\text{Area}) = 2 \pi$.
This shows the quantization of the magnetic flux penetrating a holographic vortex. 

Second, $\vartheta_3$ vanishes at
\begin{align}
  & \bm{x}_{m,n} = \left( m + \frac{1}{2} \right) \bm{b}_1
  + \left( n + \frac{1}{2} \right) \bm{b}_2 ,
\end{align}
%
for any integers $m$, $n$.
Since the expectation value of the operator $\mathcal{O}$ dual to $\Psi$
is proportional to $\gamma_L(\bmx)$,
the condensation $\langle \mathcal{O} \rangle$ has a zero
at $\bm{x}_{m,n}$.
Also, it is easily shown that the phase of
$\langle \mathcal{O} \rangle \propto \gamma_L(\bmx)$ rotates by $2\pi$
around each $\bm{x}_{m,n}$~\cite{parks}.
Thus, the cores of vortices are located at $\bm{x}_{m,n}$.

\begin{figure}
\includegraphics[width=6cm,clip]{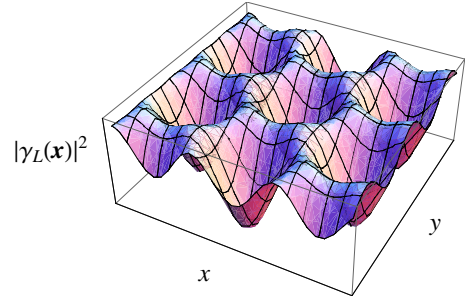}
\caption{ \label{vortex-fig}
The vortex lattice structure 
for the triangular lattice in the $(x,y)$-plane.
The vertical line represents
$\sigma = |\gamma_L|^2$
and vortex cores are located at $|\gamma_L|=0$.}
\end{figure}

The triangular lattice, where three adjoining vortices
${\bm x}_{m,n}$
form an equilateral triangle, is given by the following parameters:
\be
\label{triangle-parameters}
\frac{a_2}{a_1}=\frac{a_1}{2}=3^{-1/4}\sqrt{\pi}.
\ee
Figure~\ref{vortex-fig} shows
the configuration of
$\sigma(\bmx) = |\gamma_L(\bmx)|^2$
in the $(x,y)$-plane
for the triangular lattice.
It is well-known that this configuration minimizes
the free energy in the GL theory~\cite{parks}.

Obviously, the linear analysis presented here alone cannot lift the degeneracy of the solution and determine that the triangular solution is the correct one. To determine the correct configuration, one needs to compute the free energy and include nonlinear effects for the holographic superconductor as well, which is a goal of later sections. 

\section{Free energy and R-current}
\label{sec:free_energy}

In this section,
we calculate the free energy of the vortex lattice solution
derived in the previous section.
For simplicity, we consider the system in which the region $V$
for scalar field condensation is compact
and very large compared with the unit cell of the lattice.
So, we consider the system in which $|\Psi|$ is zero outside the region
$V$.
Since we fix both of the chemical potential $\mu=A_t(u=0)$
and the temperature $T$ under the variation of the magnetic field,
$A^{(i)}_t(u=0)=0~(i=1,2,\cdots)$ throughout this article.

\subsection{Preliminaries}

Employing the equations of motion~(\ref{eq:EOM-Psi}),
the on-shell action from Eq.~(\ref{action1}) becomes
\begin{align}
   \hS_{\text{os}}
  &= -  \frac{1}{4} \int_{\calM} d^4x~\sqrt{-g}\, F^2
  + \frac{1}{2} \int_{\partial \calM} d\Sigma_\mu \nabla^\mu | \Psi |^2
  ,
\label{onshell-action}
\end{align}
where $d\Sigma_\mu$ is the integral measure
normal to the spacetime boundary $\partial \calM$.

The boundary $\partial \calM$ in Eq.~(\ref{onshell-action}) consists of 
\begin{itemize}
\item two spacelike surfaces (the past and future surfaces), 
\item the horizon,
\item the AdS boundary $\partial \calM_1$, 
\item the boundary in the $(x,y)$-coordinates $\partial \calM_2$ ($x=\text{const.}$ and $y=\text{const.}$ surfaces).
\end{itemize}
We consider a stationary problem, so the contribution from  two spacelike surfaces cancels each other. 
There is no contribution from the horizon as well from the regularity condition at the horizon and from the ``redshift factor" $h(u \to 1) = 0$.
Thus, we need to consider the remaining two surfaces $\partial \calM_1$ and $\partial \calM_2$ to evaluate the on-shell action. 

Furthermore, one can ignore the second term in Eq.~(\ref{onshell-action}) because the scalar field $\Psi$ has a compact support in the $(x,y)$-coordinates and it satisfies the boundary condition $c_1=0$ at the AdS boundary. Therefore, the on-shell action reduces to%
\footnote{
In principle, one also should take the counterterm action $S_{\text{ct}}$ for the scalar field $\Psi$ into account, but one can easily see that $S_{\text{ct}}$ makes no contribution under the boundary condition at the AdS boundary.
}
\begin{align}
   \hS_{\text{os}}
  &= - \frac{1}{4} \int_{\calM} d^4x~\sqrt{-g}\, F^2 .
\label{eq:onshell-actionII}
\end{align}

Let us expand the on-shell action~(\ref{eq:onshell-actionII}) as
\begin{align}
\label{expansion-action}
\hS_{\text{os}} =
\hS^{(0)} + \epsilon \hS^{(1)}+\epsilon^2 \hS^{(2)}+\cdots.
\end{align}
In terms of gauge field strength $F^{(i)}_{\mu\nu}$ defined by
\begin{align}
\label{current-zeroth}
& F^{(i)}_{\mu\nu}:=\p_\mu A^{(i)}_\nu-\p_\nu A^{(i)}_\mu
& & (i=0,1,2,\cdots) ,
\end{align}
the coefficient $\hS^{(1)}$ becomes
\begin{align}
  - \hS^\scp{1}
  &= \int_{\calM} d^4x~\frac{ \sqrt{-g} }{2}\, 
    F_\scp{0}^{\mu\nu}\, F^\scp{1}_{\mu\nu}
  = \int_{\partial \calM} d\Sigma_\mu~F_\scp{0}^{\mu\nu}\, 
  A^\scp{1}_{\nu}
  ,
\label{eq:S_os-first}
\end{align}
where we used the Maxwell equations
$\nabla^\nu F^{(0)}_{\mu\nu}=0$.

Similarly, $\hS^{(2)}$ becomes
\begin{align}
   \hS^\scp{2}
  &= - \int_{\partial \calM} d\Sigma_\mu~\left(
    F_\scp{0}^{\mu\nu}\, A^\scp{2}_{\nu}
  + \frac{1}{2}\, F_\scp{1}^{\mu\nu}\, A^\scp{1}_{\nu} \right)
\nonumber \\
  &- \frac{1}{2} \int_{\calM} d^4x~\sqrt{-g}~
     j_\scp{1}^{\mu}\, A^\scp{1}_{\mu}
  ,
\label{eq:S_os-second}
\end{align}
where we used the Maxwell equations
$\nabla^\nu F^{(0)}_{\mu\nu}=0$ and
$\nabla^\nu F^\scp{1}_{\mu\nu} = j^\scp{1}_\mu$.
The last term of Eq.~(\ref{eq:S_os-second}) vanishes from the ``orthogonality condition" given in Eq.~(\ref{orthogonal-con}). Thus, the on-shell action up to second order in $\epsilon$ is given by boundary terms in Eqs.~(\ref{eq:S_os-first}) and (\ref{eq:S_os-second}).

As mentioned above, we need to consider two surfaces $\partial \calM_1$ and $\partial \calM_2$.%
 \footnote{
The surface $\partial \calM_2$ is often negligible in thermodynamic limit, but this is not the case in the presence of a magnetic field [See Eq.~(\ref{sol-A-zeroth})].
}
However, our interest is in the configuration of $\Psi$ which gives the lowest free energy, namely the $\Psi$-dependence of the free energy. Thus, it is enough to extract only the $\Psi$-dependence of the on-shell action. When $\Psi$ has a compact support in the $(x,y)$-coordinates, 
$\partial \calM_2$ gives the terms which does not depend on $\Psi$, 
so they are irrelevant.%
\footnote{
The field $\Psi$ has a compact support in the $(x, y)$-coordinates. But the gauge field $A_\mu$ has a noncompact support which depends on $\psi_1$ [See Eq.~(\ref{eq:sol-A_first})]. 
Thus, one would have $\Psi$-dependent terms from $\partial \calM_2$ through the gauge field. However, the $\psi_1$-dependent part of $A_\mu$ decays exponentially as seen in Sec.~\ref{sec:rep-G}.
}
Consequently, the boundary terms we are interested in entirely come from $\partial \calM_1$
and the on-shell action up to second order in $\epsilon$ is given by
\begin{align}
   \hS_{\text{os}}
  &= - \int_{\partial \calM_1} d\Sigma_\mu~\left(
    F_\scp{0}^{\mu\nu}\, A_{\nu} 
  + \frac{\epsilon^2}{2}\, F_\scp{1}^{\mu\nu}\, A^\scp{1}_{\nu} \right)
  + O(\epsilon^3)
\nonumber \\
  &= \left.
  \frac{\epsilon^2\, \alpha}{2} \int d^3x~\delta^{ij}
   F^\scp{1}_{u i}\, A^\scp{1}_{j}\, \right\vert_{u=0}
  + O(\epsilon^3) .
\label{onshell-action-1}
\end{align}
Note that one can ignore the first term in the first line since it does not depend on $\Psi$. Also, we used $A^{(1)}_t({\bm x}, 0)=0$.

According to the AdS/CFT dictionary,
the expectation value of the R-current
in the boundary theory, $\Exp{J_\mu(\bmx)}$
is calculated by
\begin{align}
  & \Exp{ J_\mu(t, \bmx) } = 
  \frac{L^2}{2 \kappa_4^2 e^2}\, \alpha\, F_{u \mu}(t, \bmx, u=0).
\label{eq:def-Exp_J}
\end{align}
Since $F^\scp{0}_{u i} = 0$,
\begin{align}
  & \Exp{ J_i(\bmx) }
  = \frac{L^2}{2 \kappa_4^2 e^2}\,
  \alpha\, \epsilon\, F^\scp{1}_{u i}(\bmx, u=0)
  + O(\epsilon^2) .
\label{eq:def-Exp_J_i}
\end{align}

One can rewrite the on-shell action (\ref{onshell-action-1}) by the R-current expectation value (\ref{eq:def-Exp_J_i}) as
\begin{align}
  S_{\text{os}} 
  &= \left.
  \frac{\epsilon}{2} \int d^3x~\delta^{ij}
   \Exp{J_i}\, A^\scp{1}_{j}\, \right\vert_{u=0}
  + O(\epsilon^3) .
\label{eq:onshell-action}
\end{align}
Since the free energy $\Omega$ is related to the Lorentzian on-shell action as 
$\Omega = -  S_{\text{os}}/\int dt$,
\begin{align}
   \Omega
  &= \left.
  - \frac{\epsilon}{2} \int_{\mathbb{R}^2} d\bmx~\delta^{ij}
   \Exp{J_i}\, A^\scp{1}_{j}\, \right\vert_{u=0}
  + O(\epsilon^3) .
\label{eq:def-Omega}
\end{align}
Note that these expressions ignore the terms which do not depend on $\Psi$. 

Both for the free energy and for the R-current, the problem amounts to evaluate $\Exp{J_i} \propto F^\scp{1}_{u i}$, so we obtain this quantity in the next section.

\subsection{First order solution of $A_\mu$}

The Maxwell equations at first order are given by
\begin{subequations}
\begin{align}
  & \left( \alpha^2 h\, \pdiffII{}{u} + \triangle \right) A^\scp{1}_t
  = \frac{2 L^2 \alpha^2}{u^2} A^\scp{0}_t | \psi_1 |^2,
\label{eq:EOM-At_first-pre} \\
  & \left( \pdiff{}{u}\, \alpha^2 h\, \pdiff{}{u}
  + \triangle \right) A^\scp{1}_{i}
  - \partial_i \left( \delta^{jk} \partial_j A^\scp{1}_{k} \right)
\nonumber \\
  &\hspace{0.5truecm}
  = - \frac{L^2 \alpha^2}{u^2}\, j^\scp{1}_i,
\label{eq:EOM-Ai_first-pre} \\
  & \partial_u \left( \delta^{ij} \partial_i A^\scp{1}_{j} \right)
  = 0.
\label{eq:div-Ai_first}
\end{align}
\label{eq:EOM-A_first-pre}%
\end{subequations}
We choose the gauge $A_u = 0$, but there is a residual gauge transformation $A_i \to A_i - \partial_i \Lambda(\bmx)$. From Eq.~(\ref{eq:div-Ai_first}), one can set $\delta^{ij} \partial_i A^\scp{1}_j = 0$ using the residual transformation. 
$\Exp{J_\mu} \propto F_{u \mu}$ is of course invariant under the transformation. 
In this gauge $\delta^{ij} \partial_i A^\scp{1}_j = 0$, Eq.~(\ref{eq:EOM-A_first-pre}) becomes
\begin{subequations}
\begin{align}
  & \left( \alpha^2 h\, \pdiffII{}{u} + \triangle \right) A^\scp{1}_t
  = \frac{2 \alpha^2 \rho_0^2}{u^2} A^\scp{0}_t \sigma(\bmx),
\label{eq:EOM-At_first} \\
  & \left( \pdiff{}{u}\, \alpha^2 h\, \pdiff{}{u}
  + \triangle \right) A^\scp{1}_{i}
  = \frac{\alpha^2 \rho_0^2}{u^2}\,
    \epsilon_i{}^j \partial_j \sigma(\bmx).
\label{eq:EOM-Ai_first}
\end{align}
\label{eq:EOM-A_first}%
\end{subequations}
Here, $\epsilon_{ij}$ is the antisymmetric symbol
$\epsilon_{12} = - \epsilon_{21} = 1$ and
$\epsilon_i{}^j = \epsilon_{ij}$.
We also used
\begin{align}
  & j_x^{(1)}
  = - \frac{\rho_0^2(u)}{L^2} \pdiff{\sigma(\bmx)}{y},
& & j_y^{(1)}
  = \frac{\rho_0^2(u)}{L^2} \pdiff{\sigma(\bmx)}{x},
\label{current-component}
\end{align}
from Eqs.~(\ref{eq:EOM-A}) and (\ref{def-gamma0}).

The boundary conditions for Eq.~(\ref{eq:EOM-At_first}) are
$0 = A^\scp{1}_t(u=1) = A^\scp{1}_t(u=0)$. 
For Eq.~(\ref{eq:EOM-Ai_first}), we impose the boundary conditions such that the solution is regular at the horizon and $\epsilon F^\scp{1}_{xy} = B - B_{c2}$ at the AdS boundary, namely,  $2 \partial_{[x} A_{y]}^\scp{1}(u=0) = - B_{c2}$.

The solutions are formally written by introducing Green functions which satisfy the Dirichlet condition at the AdS boundary:
\begin{subequations}
\begin{align}
   A^\scp{1}_t
  &= - 2 \alpha^2 \int^1_0 du'~
  \frac{\rho_0^2(u')}{u'^2 h(u')} A^\scp{0}_t(u')
\nonumber \\
  &\times
  \int d\bmx' G_t(u;u'|\, \bmx - \bmx' )\, \sigma(\bmx'),
\label{eq:sol-At_first} \\
   A^\scp{1}_i
  &= a_i(\bmx)
  - \alpha^2 \epsilon_i{}^j \int^1_0 du'~\frac{\rho_0^2(u')}{u'^2}
\nonumber \\
  &\times
  \int d\bmx' G_B(u;u'|\, \bmx - \bmx' )\, \partial_j \sigma(\bmx').
\label{eq:sol-Ai_first}
\end{align}
\label{eq:sol-A_first}%
\end{subequations}
Here, $a_i(\bmx)$ is a homogeneous solution of Eq.~(\ref{eq:EOM-Ai_first}) satisfying $2 \partial_{[x} a_{y]} = - B_{c2}$ at the AdS boundary, and is independent of $u$.
$G_t$ and $G_B$ are Green functions of Eqs.~(\ref{eq:EOM-At_first}) and (\ref{eq:EOM-Ai_first}), respectively: 
\begin{subequations}
\begin{align}
  & \left( \alpha^2\, h\, \pdiffII{}{u} + \triangle \right)
    G_t(u;u'| \bmx)
  = - h(u)\, \delta(u-u') \delta(\bmx),
\label{eq:EOM-G_t} \\
  & G_t(u=0;u'| \bmx) = G_t(u=1;u'| \bmx) = 0,
\label{eq:bc-G_t}
\end{align}
\label{green-t}%
\end{subequations}
\begin{subequations}
\begin{align}
  & \left[ \alpha^2 \pdiff{}{u} \left( h\, \pdiff{}{u} \right)
  + \triangle \right] G_B(u;u'| \bmx)
  = - \delta(u-u') \delta(\bmx),
\label{eq:EOM-G_B} \\
  & G_B(u=0;u'| \bmx)
  = \lim_{u \to 1} h(u) \partial_u G_B(u;u'| \bmx) = 0.
\label{eq:bc-G_B}
\end{align}
\label{green-B}%
\end{subequations}
The boundary condition Eq.~(\ref{eq:bc-G_B}) at $u = 1$ follows from the regularity at the horizon. Similar expressions hold for $u \leftrightarrow u'$ from $G(u;u')=G(u';u)$.

\subsection{Free energy and R-current}

Using the Green function $G_B$, one can express the R-current expectation value as
\begin{align}
   \Exp{J_i(\bmx)}
  &= - \epsilon_i{}^j \partial_j \Xi(\bmx) ,
\label{eq:Exp_J_i}
\end{align}
where
\begin{align}
   \Xi(\bmx)
  &:= \epsilon\, \frac{L^2 \alpha^3}{2 \kappa_4^2 e^2}
  \int_{\mathbb{R}^2} d\bmx' \sigma(\bmx')
\nonumber \\
  &\times \left.
  \partial_u \int^1_0 du' \frac{\rho_0^2(u')}{u'^2}
  G_B(u ; u' | \bmx - \bmx' ) \right\vert_{u=0} .
\label{eq:def-Xi}
\end{align}
Here, we used $\partial G_B/\partial x'^i = - \partial G_B/\partial x^i$.
Equation~(\ref{eq:Exp_J_i}) implies a circulating R-current which is dictated simply from the current conservation. It is a different issue though whether the current actually flows 
along the lines of constant field for the condensate 
like the GL theory. 

The free energy can be expressed by the Green function similarly. The free energy is given by 
\begin{align*}
   \Omega
  &= \left.
  \frac{\epsilon}{2} \int_{\mathbb{R}^2} d\bmx~
    \epsilon^{jk} \left( \partial_k \Xi \right)
   A^\scp{1}_{j}\, \right\vert_{u=0}
  + O(\epsilon^3)
\nonumber \\
  &= - \frac{\epsilon B_{c2}}{2} \int_{\mathbb{R}^2} d\bmx~\Xi
  + O(\epsilon^3)
  ,
\end{align*}
using Eq.~(\ref{eq:def-Omega}) and $2 \partial_{[x} A_{y]}^\scp{1}(u=0) = - B_{c2}$. 
From Eqs.~(\ref{green-B}), $G_B$ satisfies
%
\begin{align*}
  & \alpha^2 \odiff{}{u} \left( h\, \odiff{}{u} \right)
    \int_{\mathbb{R}^2} d\bmx\, G_B(u;u'| \bmx)
  = - \delta(u-u'),
\end{align*}
%
and
\begin{align}
  & \alpha^2 \int_{\mathbb{R}^2} d\bmx\, G_B(u;u'| \bmx)
  = \int^{\text{min}(u, u')}_0 \frac{du''}{h(u'')}~.
\label{eq:int-G_B}
\end{align}
Consequently, one obtains
\begin{align}
  & \Omega
  = - \frac{L^2}{2 \kappa_4^2 e^2}\,
  \frac{\epsilon^2 B_{c2}\, \alpha}{2}~C
  \, \text{vol}(V)\, \overline{\sigma}~,
\label{free-energy1}
\end{align}
where
\be
\label{coefficient}
C:=\int^1_0 du~\frac{\rho_0^2(u)}{u^2} > 0,
\ee
and $\overline{f}$ indicates the average of $f$ over $V$ in the $(x,y)$-plane:
\begin{align}
  & \overline{f}
  := \frac{1}{\text{vol}(V)}\, \int_{V} d\bmx~f(\bmx) .
\label{eq:def-overline}
\end{align}

From Eq.~(\ref{free-energy1}), $\Omega<0$, which suggests that the free energy of 
the vortex lattice state is smaller than the one of the normal state
 ($\Psi = 0$).

Equations~(\ref{eq:Exp_J_i}), (\ref{eq:def-Xi}) and (\ref{free-energy1})
are not our final results.
Because $\rho_0$ and $\gamma_L$
are the solutions of the linear equations
(\ref{eq-rho}) and (\ref{eq-gamma}),
their normalizations have not been determined yet.
Below we eliminate this ambiguity, so $\Exp{J_\mu}$ and $\Omega$ are characterized only
by two lattice parameters $a_1$ and $a_2$.


The ambiguity in the normalizations come from the linear equations, and it is resolved only after one considers nonlinearity. So, let us consider the equation of motion for $\psi_2$.
As shown in Appendix A, if $\psi_2$ obeys the boundary condition
$c_1=0$ and the regularity condition
at the horizon, we obtain the ``orthogonality condition" 
(\ref{orthogonal-con}) which relates $\rho_0$ to the gauge field $A_\mu^{(1)}$.

Using Eqs.~(\ref{current-component}) and integration by parts,
the ``orthogonality condition" (\ref{orthogonal-con}) 
can be written as
\begin{align}
\label{orthogonal-con1}
& \int^1_0du~\frac{\rho_0^2(u)}{u^2}
\int_{\mathbb{R}^2} d\bm{x}~F^\scp{1}_{xy}\, \sigma(\bmx)
\nonumber \\
& =2\int^1_0du~\frac{\rho_0^2(u)}{u^2 h(u)}\, A^{(0)}_t(u)
\int_{\mathbb{R}^2} d\bm{x}~A^{(1)}_t\, \sigma(\bmx) .
\end{align}
From Eq.~(\ref{eq:sol-Ai_first}),
\begin{align}
   F^\scp{1}_{xy}
  &= - B_{c2}
  + \alpha^2  \int^1_0 du'\frac{\rho_0^2(u')}{u'^2}
   \triangle \big[\, G_B(u;u') * \sigma\, \big](\bmx)
   ,
\label{sol-green-B1}
\end{align}
where \lq\lq\, $*$\, \rq\rq~is convolution
in the $(x,y)$-plane:
\begin{align}
  & \big[\, f * g\, \big](\bmx)
  := \int d\bmx' f( \bmx - \bmx' )\, g(\bmx').
\label{eq:def-convolution}
\end{align}
Substituting the solutions (\ref{eq:sol-At_first}) and 
(\ref{sol-green-B1}) into Eq.~(\ref{orthogonal-con1}),
we obtain
\begin{align}
   \frac{B_{c2} C}{\overline{\sigma}}
  &= \alpha^2 \int^1_0 du\, \frac{\rho_0^2(u)}{u^2}
  \int^1_0 du'\, \frac{\rho_0^2(u')}{u'^2}\, \calI(u, u') ,
\label{eq:coefficient-C-pre}
\end{align}
where we define
\begin{align}
   \calI
  &:= \frac{2 A^\scp{0}_t(u)}{h(u)}~
  \frac{ \overline{ \sigma\, \big[\, G_t(u;u') * \sigma\, \big] } }
       {(\overline{\sigma})^2}~
  \frac{2 A^\scp{0}_t(u')}{h(u')}
\nonumber \\
  &+ \frac{ \overline{ \sigma\, \triangle
              \big[\, G_B(u;u') * \sigma\, \big] } }
          {(\overline{\sigma})^2} ,
\label{eq:def-calI}
\end{align}
which is independent of the normalization of $\gamma_L$.

Then, the free energy (\ref{free-energy1}) is expressed by 
\begin{align}
  & \Omega
  = - \frac{L^2}{2 \kappa_4^2 e^2}\,
  \frac{\epsilon^2 B_{c2}^2\, \alpha}{2}\,
  \frac{ \text{vol}(V) }{\Gamma} ,
\label{eq:Omega}
\end{align}
where 
\begin{align}
   \Gamma
  &:= \frac{B_{c2}}{C\, \overline{\sigma}}
  = \frac{\alpha^2}{C^2}
  \int^1_0 du\, \frac{\rho_0^2(u)}{u^2}
  \int^1_0 du'\, \frac{\rho_0^2(u')}{u'^2}\, \calI(u, u')
\nonumber \\
  &= \alpha^2 \int^1_0 du\, \frac{\hrho_0^2(u)}{u^2}
  \int^1_0 du'\, \frac{\hrho_0^2(u')}{u'^2}\, \calI(u, u'). 
\label{eq:def-Gamma}
\end{align}
Here, $\hrho_0$ is the solution of Eq.~(\ref{eq-rho}) normalized by
\be
\int^1_0 du~\frac{\hat{\rho}_0^2(u)}{u^2} =1. 
\ee
Since the functions $\calI$ and $\Gamma$ do not depend on the 
normalization of $\rho_0$ and $\gamma_L$, 
the free energy (\ref{eq:Omega}) also does not depend on the normalization either.

Similarly, the potential $\Xi$ (\ref{eq:def-Xi}) which gives the R-current expectation value does not have the ambiguity in normalization. This can be seen by expressing the potential as
\begin{align}
   \Xi(\bmx)
  &= \epsilon\, B_{c2}\, \frac{L^2 \alpha^3}{2 \kappa_4^2 e^2}\,
  \frac{1}{\Gamma}\, \partial_u
\nonumber \\
  &\times \left.
  \int^1_0 du' \frac{\hrho_0^2(u')}{u'^2}\,
  \frac{ \big[\, G_B(u ; u' ) * \sigma\, \big](\bmx) }
       {\overline{\sigma}}\, \right\vert_{u=0} .
\label{eq:Xi}
\end{align}

This is our main result: the free energy (\ref{eq:Omega}) and
the expectation value of the R-current (\ref{eq:Exp_J_i}) are expressed by $\Gamma$ (\ref{eq:def-Gamma}) and $\Xi$ (\ref{eq:Xi}).

As seen in these expressions, 
the free energy density and the R-current are not only determined
by the complex scalar field $\sigma = |\gamma_L|^2$
at $\bmx$
but also by $\sigma$ in the entire region around $\bmx$.
On the other hand, in the GL theory, the free energy density is a local function of the order parameter $\Psi_{\text{GL}}(\bmx)$. The superconducting current is also expressed locally by 
\begin{align}
  & J^{\text{GL}}_i(\bmx)
  = - \epsilon_i{}^j \partial_j |\Psi_{\text{GL}}(\bmx)|^2.
\label{GL-current}
\end{align}
This implies that the superconducting current
flows along the lines of 
$|\Psi_{\text{GL}}|^2=$ constant~\cite{parks}.
These differences are natural since we have not taken a long-wavelength limit in evaluating the free energy and the R-current unlike the GL theory. 
By taking a long-wavelength limit, our result should reduce to the GL theory. 
As a result, the triangular lattice solution, which is the most favorable solution in the GL theory, should also become the most favorable one in the holographic superconductor. One should reproduce Eq.~(\ref{GL-current}) as well. We will demonstrate this in the next section.

\section{The long-wavelength limit and the triangular lattice}
\label{sec:low_energy_limit}

\subsection{A representation of Green functions}
\label{sec:rep-G}

The expression of $\Gamma$ in Eq.~(\ref{eq:def-Gamma}) is nonlocal both in the AdS radius direction and in the $(x,y)$-directions, which makes the boundary interpretation rather unclear. The nonlocalities come from Green functions $G_B$ and $G_t$, so it is useful to expand the $u$-dependence of the Green functions by a complete set of orthonormal functions. 

First, let us consider eigenfunctions $\chi_\lambda(u)$
satisfying
\begin{subequations}
\begin{align}
  & \calL_B \chi_\lambda(u) = \lambda \chi_\lambda(u),
\hspace{0.5truecm}
   \calL_B
  := - \alpha^2 \odiff{}{u} \left ( h \odiff{}{u} \right),
\label{eq:eigen_eqn-B} \\
  & \chi_\lambda(u=0) = \lim_{u \to 1} h(u) \chi_\lambda'(u) = 0.
\label{eq:eigen_bc-B}
\end{align}
\label{eq:eigen_sys-B}%
\end{subequations}
%
The operator $\calL_B$ becomes an Hermitian operator
for the inner product defined by
\begin{align}
  & \IP{\phi}{\psi}_B
  := \int^1_0 du~\phi^\dagger(u)\, \psi(u) ,
\label{eq:def-IP_B}
\end{align}
for any solutions $\phi$ and $\psi$ of Eq.~(\ref{eq:eigen_sys-B}).
One can easily show $\lambda > 0$.

The normalized eigenfunctions $\{ \chi_\lambda \}$
with respect to the inner product (\ref{eq:def-IP_B})
form a complete orthonormal set
\begin{align}
  & \langle \chi_\lambda \vert \chi_{\lambda'} \rangle_B
  = \delta_{\lambda\lambda'},
& & \sum_\lambda  \chi_\lambda(u) \chi_\lambda^\dagger(u')
  = \delta(u-u'),
\label{orthonormal-set1}
\end{align}
and the Green function $G_B$ in Eq.~(\ref{green-B}) is represented
by $\{ \chi_\lambda \}$ as
\begin{align}
  & G_B(u;u'| \bmx)
  = \sum_{\lambda>0} \chi_\lambda(u) \chi_\lambda^\dagger(u')\,
    G_2( \bmx ; \lambda).
\label{sol-green-B}
\end{align}
Here, $G_2( \bmx ; \mathfrak{m}^2)$ is the Green function defined
on $(x,y)$-plane as
\begin{align}
  & ( \triangle - \mathfrak{m}^2 ) G_2( \bmx ; \mathfrak{m}^2) = - \delta(\bmx) .
\label{green-two}
\end{align}
The solution of Eq.~(\ref{green-two}) satisfying $\lim_{|{\bm x}|\to\infty }|G_2|<\infty$ is given
by the modified Bessel function:
\be
\label{sol-green-two}
G_2( \bm{x}; \mathfrak{m}^2)
= \frac{1}{2 \pi} K_0(\mathfrak{m} |\bmx |)
\ee
for any real positive value of $\mathfrak{m}^2$.

Similarly, $G_t$ in Eq.~(\ref{green-t}) is constructed as
\begin{align}
  & G_t(u;u'| \bmx )
  = \sum_{\eta>0} \xi_\eta(u) \xi_\eta^\dagger(u')\,
    G_2( \bmx ; \eta) ,
\label{sol-green-t}
\end{align}
where $\{ \xi_\eta \}$ is the complete orthonormal eigensystem
of the equation
\begin{subequations}
\begin{align}
  & \calL_t\, \xi_\eta(u) = \eta\, \xi_\eta(u),
\hspace{0.8truecm}
   \calL_t
  := - \alpha^2 h\, \odiffII{}{u},
\label{eq:eigen_eqn-t} \\
  & \xi_\eta(u=0) = \xi_\eta(u=1) = 0,
\label{eq:eigen_bc-t}
\end{align}
\label{eq:eigen_sys-t}%
\end{subequations}
with
\begin{align}
  & \langle \xi_\eta | \xi_{\eta'} \rangle_t
  = \delta_{\eta\eta'},
& & \sum_\eta \xi_\eta(u) \xi_\eta^\dagger(u') 
  = h(u) \delta(u-u').
\label{orthonormal-set2}
\end{align}
The inner product $\IP{}{}_t$ is defined by
\begin{align}
  & \IP{\phi}{\psi}_t
  := \int^1_0 \frac{du}{h(u)}~\phi^\dagger(u)\, \psi(u)
\label{eq:def-IP_t}
\end{align}
for any solutions $\phi$ and $\psi$ of Eq.~(\ref{eq:eigen_sys-t}).
The operator $\calL_t$ is Hermitian with respect to
the inner product (\ref{eq:def-IP_t}),
and $\eta>0$.

Substituting Eqs.~(\ref{sol-green-B}) and (\ref{sol-green-t})
into Eq.~(\ref{eq:def-calI}),
we obtain another expression of $\Gamma$ as
\begin{align}
  & \Gamma
  = \sum_\lambda P(\lambda) \zeta_1(\lambda)
  + \sum_\eta Q(\eta) \zeta_0(\eta),
\label{Gamma}
\end{align}
where we introduce $\zeta_n(\mathfrak{m}^2)$
\begin{align}
   \zeta_n(\mathfrak{m}^2)
  &:= \frac{ \overline{ \sigma\,
                 \triangle^n \big[\, G_2(\mathfrak{m}^2) * \sigma\, \big] } }
             { (\overline{\sigma})^2 },
\label{eq:def-zeta_n}
\end{align}
%
and the functions $P$, $Q$ defined by
\begin{align}
  & P(\lambda)
  := \left| \int^1_0 du~\chi_\lambda^\dagger\, \frac{\hrho_0^2}{u^2}\,
  \right|^2,
\label{eq:def-P} \\
  & Q(\eta)
  := \left| \int^1_0 \frac{du}{h}~\xi_\eta^\dagger\,
  (2 A_t^\scp{0})\, \frac{\hrho_0^2}{u^2}\, \right|^2.
  \label{eq:def-Q}
\end{align}
Similarly, $\Xi$ (\ref{eq:Xi}) is represented as
\begin{align}
  & \Xi(\bmx)
  = \epsilon\, B_{c2}\, \frac{L^2 \alpha^3}{2 \kappa_4^2 e^2}\,
  \frac{1}{\Gamma}\,
\nonumber \\
  &\times \sum_{\lambda} \chi'_\lambda(0)
  \left( \int^1_0 du~\chi^\dagger_{\lambda}\, \frac{\hrho_0^2}{u^2}
  \right)
  \frac{ \big[\, G_2(\lambda) * \sigma\, \big](\bmx) }
       {\overline{\sigma}}
  .
\label{eq:Xi-II}
\end{align}
The eigenvalues $\lambda$ and $\eta$ can be obtained numerically
by solving the two differential equations~(\ref{eq:eigen_sys-B})
and (\ref{eq:eigen_sys-t}).
Both are positive, and the minimum values of $\lambda$ and $\eta$
are given by $\lambda\simeq2.22\alpha^2$ and $\eta\simeq7.41\alpha^2$,
respectively.

The holographic superconductor
is constructed in the gravity theory with one extra dimension
which is extended perpendicular to the $(2+1)$-dimensional spacetime
of the boundary theory.
So, from the $(2+1)$-dimensional point of view,
the bulk gauge fields $A_\mu$ 
with a wide variety of mass $\sqrt{\lambda}$ and $\sqrt{\eta}$ appear
as in Eqs.~(\ref{eq:eigen_sys-B}) and (\ref{eq:eigen_sys-t}).

\subsection{The long-wavelength limit}

We find that both the free energy density and the R-current take nonlocal forms whereas they take local forms in the GL theory. This is because the AdS/CFT results correspond to the results to all orders in effective theory expansion. The GL theory takes only first few terms in the effective theory expansion, so one needs to take a long-wavelength limit in the AdS/CFT results to compare with the GL theory. Taking such a limit is common in hydrodynamic studies based on the AdS/CFT duality (See, e.g., Refs.~\cite{Natsuume:2007qq,Son:2007vk,Natsuume:2008ha}.)

In this subsection, we demonstrate that our results indeed reduce to the GL theory ones in the long-wavelength limit. As a result, the triangular lattice solution becomes the most favorable one like the GL theory. 

We have obtained nonlocal expressions, but there is in fact an analogous situation in a superconductor. A superconductor has small length scales, the Pippard/BCS coherence length and the mean-free path. In the presence of these small scales, the electromagnetic response in general takes a nonlocal form. The local form such as the London equation is the limit where these length scales are negligible \cite{parks}.

In our problem, the nonlocalities come from the convolution in Eqs.~(\ref{eq:def-zeta_n}) and (\ref{eq:Xi-II}), i.e.,  
\be
\big[\, G_2 (\lambda) * \sigma\, \big](\bmx)
 = \int d\bmx' G_2( \bmx - \bmx'; \lambda)\, \sigma(\bmx') 
\ee
and a similar expression for $G_2(\eta)$. 
The Green's functions $G_2$ have the natural length scales $1/\sqrt{\lambda}$ and $1/\sqrt{\eta}$. They are the small length scales in our problem. Their microscopic interpretation is unclear but they are $O(T^{-1})$. On the other hand, the natural length scale of the vortex lattice is the size of the fundamental region parametrized by $r_0$. This is the length scale of the condensate $\sigma = |\gamma_L|^2$. When $r_0 \gg 1/\sqrt{\lambda}, 1/\sqrt{\eta}$, $G_2$ quickly decays compared with $\sigma$, and the convolution reduces to a local form. This is what we meant by the ``long-wavelength limit."

By replacing $\sigma(\bmx'\,)$ by $\sigma(\bmx)$ in the convolution, we obtain
\begin{align}
  & \big[\, G_2(\lambda) * \sigma\, \big](\bmx)
  = \frac{\sigma(\bmx)}{\lambda}
  \left[~1 + O\left( \frac{1}{r_0^2 \lambda} \right)~\right]~,
\label{eq:approx-convolution}
\end{align}
where we used
\begin{align*}
  & \int_{\mathbb{R}^2} d\bmx~G_2(\bmx ; \lambda) = \frac{1}{\lambda}.
\end{align*}
Subleading terms in Eq.~(\ref{eq:approx-convolution}) can be estimated from a series expansion $\sigma(\bm{x}') = \sigma(\bm{x}) + \cdots$.

Substituting Eq.~(\ref{eq:approx-convolution}) into Eq.~(\ref{eq:def-zeta_n}), one obtains the coefficient $\Gamma$~(\ref{Gamma}) as
\begin{align}
  & \Gamma
  = C_1 \frac{ \overline{\sigma^2} }{ (\overline{\sigma})^2 }
  \left[~1 + O\left( \frac{1}{r_0^2 \eta},
                     \frac{1}{r_0^2 \lambda} \right)~\right]~,
\label{eq:local-Gamma}
\end{align} 
where
\begin{align}
   C_1
  &:= \sum_{\eta>0} \frac{Q(\eta)}{\eta} .
\end{align}
$\Gamma > 0$ from its definition (\ref{eq:def-Gamma}). [The constant $C_1$ is indeed positive from Eq.~(\ref{eq:def-Q}) and the positivity of $\eta$.]
One has the lowest free energy $\Omega$  when $\Gamma$ takes its minimum. 
Therefore, the thermodynamically realized configuration is given when 
\begin{align}
\beta:=\frac{\overline{\sigma^2}}{(\overline{\sigma})^2}
\end{align}
takes the minimum value. 
This is the same condition as the one for the Abrikosov lattice in standard type II superconductors \cite{parks}. As is well-known, the minimum is $\beta\simeq1.16$, which is given by the triangular lattice (\ref{triangle-parameters}).

Similarly, the R-current (\ref{eq:Xi-II}) becomes
\begin{align}
   \Xi(\bmx)
  &= \epsilon\, B_{c2}\, \frac{L^2 \alpha}{2 \kappa_4^2 e^2}\,
  \frac{1}{\Gamma}\, \frac{\sigma(\bmx)}{\overline{\sigma}}
  \left[~1 + O\left( \frac{1}{r_0^2 \lambda} \right)~\right]~,
\label{eq:approx-Xi}
\end{align}
using Eq.~(\ref{eq:approx-convolution}).
Thus, like the GL theory, the circulating R-current flows 
along the lines of constant field for the condensate.

Let us consider the validity of our approximation. One can achieve the long-wavelength limit by taking a small magnetic field $B_{c2}$
since $r_0 = 1/\sqrt{B_{c2}}$. 
As shown in Fig.~\ref{fig:phase_diagram}, one can take an arbitrary small value of $B_{c2}$ by choosing the temperature $T$ suitably.
Using $q=\mu/\alpha, \lambda \sim 2\alpha^2$, and $q_c \sim 4$, the condition $B_{c2} \ll \lambda, \eta$ gives
\begin{equation}
\frac{B_{c2}}{\mu^2} 
\ll \frac{2}{q_c^2} \left(\frac{T}{T_c}\right)^2 
\sim \frac{1}{8} \left(\frac{T}{T_c}\right)^2~.
\end{equation}
This condition is implemented in Fig.~\ref{fig:phase_diagram}. Our approximation in this subsection becomes good if the system is located well below the dashed curve in the phase diagram.

Incidentally, one would incorrectly conclude that
the expanded solution of Eqs.~(\ref{expansion}) exists
even for $B>B_{c2}$ by replacing the deviation parameter
with $\epsilon=(B-B_{c2})/B_{c2}~(>0)$.
Suppose that this is the case.
Then, this changes the sign in the first term of $F^\scp{1}_{xy}$
in the solution~(\ref{sol-green-B1}).
As a result, one has an extra minus sign in the following equations:
the left-hand side of Eq.~(\ref{eq:coefficient-C-pre})
and the right-hand side of the second line of Eq.~(\ref{eq:def-Gamma}).
But this contradicts with the result in Eq.~(\ref{eq:local-Gamma}). 
Therefore, the $\epsilon$-expansion of $\Psi$ does not exist
for $B>B_{c2}$ at least
when $r_0 \gg 1/\sqrt{\lambda},1/\sqrt{\eta}$.
This agrees with the argument in Sec.~II
that  the superconducting phase ceases to exist at $B=B_{c2}$
and the normal phase appears for $B>B_{c2}$.

\section{Conclusions and discussion}\label{sec:discussion}

We have investigated the vortex lattice solution
in the holographic superconductor.
One would interpret the holographic superconductor as a superfluid, and the computations described here can equally apply to 
this case as well in a slight modification. In this case, the rotation of the superfluid is analogous to the magnetic field. 

One main difference between the conventional superconductors
in the GL theory and the holographic superconductor is
that both the free energy density and the R-current are expressed
by nonlocal quantities, such as the two-point function $G_2$
in Eq.~(\ref{green-two}). 
This is because our results in Sec.~\ref{sec:free_energy} is beyond the applicability of the GL theory. In fact, we show in Sec.~\ref{sec:low_energy_limit} that our results reduce to the results of the GL theory by taking a long-wavelength limit. Then, one should be able to estimate the corrections to the GL theory by inspecting our expressions in Sec.~\ref{sec:free_energy}.  


Another interesting direction to pursue is
the investigation of the kinetics of the vortex lattice solution.
For example, it is well-known that vortices move
at a constant velocity along the direction
perpendicular to the magnetic field by the Lorentz force
and the electric resistance appears
if we add an appropriate external electric field
perpendicular to the magnetic field.
This phenomena is important for pinning of superconductors
with irregularities~\cite{parks}.
We will discuss this in more detail in Ref.~\cite{mno}.
It is also interesting to consider the dynamic critical phenomena
of the vortex lattice solution,
as investigated in Refs.~\cite{mno2008,mno2009}.

\begin{acknowledgments} 
We would like to thank Elena Caceres, Gary Horowitz, Esko Keski-Vakkuri, Clifford Johnson, and Sean Nowling
for useful discussions.
MN would also like to thank the Aspen Center for Physics for their hospitality and for a stimulating environment while part of this work was carried out.
This research was supported in part by the Grant-in-Aid for Scientific
Research~(20540285) from the Ministry of Education, Culture,
Sports, Science and Technology, Japan.
\end{acknowledgments}

\appendix
\section{``Orthogonality condition"}

The equations for $\psi_1$ and $\psi_2$ are given by
%
\begin{align}
  & \left( D_\scp{0}^2 - m^2 \right) \psi_1 = 0~,
\hspace{0.5truecm}
    \left( D_\scp{0}^2 - m^2 \right) \psi_2 = J~,
\label{eq:EOM-psi2} \\
  & J
  := i \left\{ D^\scp{0}_\mu \left( A^\mu_\scp{1}\, \psi_1 \right)
  + A^\mu_\scp{1} D^\scp{0}_\mu \psi_1 \right\}~,
\label{eq:def-J}
\end{align}
%
where $D_\scp{0}^\mu := \nabla^\mu - i A^\mu_\scp{0}$.

Recall that the field $\Psi$ has a compact support in the $(x, y)$-coordinates and satisfies the regularity condition at the horizon and $c_1=0$ at the AdS boundary. 
Then, using Eq.~(\ref{eq:EOM-psi2}) and integration by parts, one obtains the ``orthogonality condition:"
\begin{align}
   0
  &= \int_{\calM} d^4x~\sqrt{-g}\, \left\{ \psi_1^\dagger 
       \left( D_\scp{0}^2 - m^2 \right) \psi_2
     - \psi_1^\dagger\, J \right\}
\nonumber \\
  &= \int_{\partial\calM} d\Sigma_\mu
     \left\{ \psi_1^\dagger D^\mu_\scp{0} \psi_2
     - \left( D^\mu_\scp{0} \psi_1 \right)^\dagger \psi_2
     - i\, A^\mu_\scp{1}\, \vert \psi_1 \vert^2 \right\}
\nonumber \\
  &+ \int_{\calM} d^4x~\sqrt{-g}\, A^\mu_\scp{1}\, j^\scp{1}_\mu
\nonumber \\
  &= \int_{\calM} d^4x~\sqrt{-g}\, A^\mu_\scp{1}\, j^\scp{1}_\mu~.
\label{orthogonal-con}
\end{align}

\end{document}